\documentclass[aps,prb,preprint,groupedaddress]{revtex4}

\usepackage{graphicx}
\usepackage{dcolumn}
\newcommand{\tn}{\textnormal}

\begin{document}


\title{CeRuPO: A rare example of a ferromagnetic Kondo lattice}

\author{C. Krellner}
\email{krellner@cpfs.mpg.de}
\author{N.S. Kini}
\altaffiliation{Present address: Chemistry and Physics of Materials Unit, Jawaharlal Nehru Centre for Advanced Scientific Research, Jakkur P.O., Bangalore-560064, India}
\author{E.M. Br\"uning}
\author{K. Koch}
\author{H. Rosner}
\author{M. Nicklas}
\author{M. Baenitz}
\author{C. Geibel}
\affiliation{Max Planck Institute for Chemical Physics of Solids, N\"othnitzer Str. 40, D-01187 Dresden, Germany }

\date{\today}

\begin{abstract}
We have determined the physical ground state properties of the compounds CeRuPO and CeOsPO by means of magnetic susceptibility $\chi(T)$, specific heat $C(T)$, electrical resistivity $\rho(T)$, and thermopower $S(T)$ measurements. $\chi(T)$ reveals a trivalent $4f^1$ cerium state in both compounds. For CeRuPO a pronounced decrease of $\rho(T)$ below 50\,K indicates the onset of coherent Kondo scattering, which is confirmed by enhanced $S(T)$. The temperature and magnetic field dependence of $\chi(T)$ and $C(T)$ evidence ferromagnetic (FM) order at $T_{\rm C}=15$\,K. Thus, CeRuPO seems to be one of the rare example of a FM Kondo lattice. In contrast, CeOsPO shows antiferromagnetic order at $T_{\rm N}=4.4$\,K despite only minor changes in lattice parameters and electronic configuration. Additional $^{31}$P NMR results support these scenarios. LSDA+$U$ calculations evidence a quasi two dimensional electronic band structure, reflecting a strong covalent bonding within the CeO and RuP layers and a weak ionic like bonding between the layers.
\end{abstract}


\maketitle

\section{\label{Intro}Introduction}
Intermetallic Kondo lattice systems have attracted considerable attention in the last decades. In these systems, two different interactions determine the ground state properties, which can be understood in a simple way in terms of the Doniach picture. \cite{Doniach:1977} The formation of a nonmagnetic Kondo singlet competes with the formation of a magnetic ordered ground state; \cite{Steglich:1996} in the case of dominating Kondo interactions, the magnetic moments of the $4f$ electrons are screened by the conduction electrons, and a nonmagnetic ground state develops. If the Ruderman-Kittel-Kasuya-Yosida (RKKY) exchange interaction exceeds the Kondo screening, long range magnetic order occurs. While many Ce-based Kondo lattices show antiferromagnetic (AFM) ground states, only very few systems are known with ferromagnetic (FM) order and pronounced Kondo effects. Therefore, the behavior at a FM critical point in a Kondo lattice is presently not settled. \cite{Kirkpatrick:2003} From the few FM Kondo systems known, the best studied examples are YbNiSn, \cite{Drescher:1996} CeSi$_{2-x}$, \cite{Drotziger:2006} and the compound-series CePd$_{1-x}$Rh$_x$. \cite{Sereni:2007} However, in the former compound pressure stabilizes the magnetic state, preventing a study of the critical region. In the latter, disorder is not negligible and may hide the intrinsic physical properties. On the way to look for new Ce-based Kondo lattice systems close to a FM quantum phase transition, the CeTPO (T\,=\,transition metal) compound series attracted our interest, because of the rather unusual crystal structure with alternating layers of TP$_4$ and OCe$_4$ tetrahedra, causing Ce-Ce inter-atomic distances to be in the range where FM order can be observed. \cite{Sereni:2006a} This compound series, crystallizing with the tetragonal ZrCuSiAs type structure, was discovered by the Jeitschko group; \cite{Zimmer:1995, Kaczorowski:1994} but no physical properties have been reported for CeRuPO. The compound CeOsPO has not been synthesized before. Recently, LaFePO was found to be superconducting, supporting the interest in this class of material. \cite{Kamihara:2006,Lebegue:2007}

In this contribution, we will present the low temperature physical properties of CeRuPO and CeOsPO. The article is organized as follows: In section \ref{Prep} we will discuss the details of the sample preparation and the crystal structure. The physical properties of CeRuPO are presented in detail in section \ref{CeRuPO} and compared with the results of CeOsPO in section \ref{CeOsPO}. We present and discuss $^{31}$P NMR results for both compounds in section \ref{NMR}. Finally, in section \ref{Theory} we show band structure calculations of both compounds using an LSDA+$U$ approach and analyze the bonding properties and the magnetic exchange.

\section{\label{ExpDet}Experimental}
\subsection{\label{Prep}Preparation and Crystal structure}
The polycrystalline samples were prepared using a Sn-flux method in evacuated quartz tubes. \cite{Kanatzidis:2005} For CeRuPO the preparation was optimized with RuO$_2$ as the source for the oxygen. We found the best growing conditions for an initial molar ratio of 8:6:4:82 (Ce:P:RuO$_2$:Sn). This preparation route was also used to obtain the LaRuPO compound, serving as the nonmagnetic reference. CeOsPO was prepared with SnO$_2$ instead of the hazardous osmiumtetraoxide. For this compound we found the best results for the initial molar ratio of 4:3:2:2:89 (Ce:Os:P:SnO$_2$:Sn). The heating schedule was similar for all preparation cycles. The starting mixture was heated with a rate of 100$^{\circ}$C per hour to 1000$^{\circ}$C and kept at this temperature for 5 hours; subsequently, the melt was cooled down to 500$^{\circ}$C by moving the quartz tube out of the furnace with an average cooling rate of 5$^{\circ}$C per hour. After the reaction the excess Sn was dissolved in dilute HCl which attacks the CeTPO at a lower rate. Small single crystals ($<50$\,$\mu$m) of CeTPO are attached to each other forming flake-like polycrystals which were ground and pelletized for further physical characterization. Several samples were investigated with energy dispersive X-ray analysis. These measurements reveal a stoichiometric Ce:Ru:P content and confirmed the presence of oxygen in the compounds; furthermore, no foreign phases could be detected. In addition, carrier gas-hot extraction (LECO, TCH 600) was used to determine the oxygen content $x_{\rm O}$ of CeRuPO more precisely. The result of $x_{\rm O}=(25.0\pm0.7)$\,at\% indicates a stoichiometric oxygen occupancy.

\begin{figure}
\includegraphics[width=10cm]{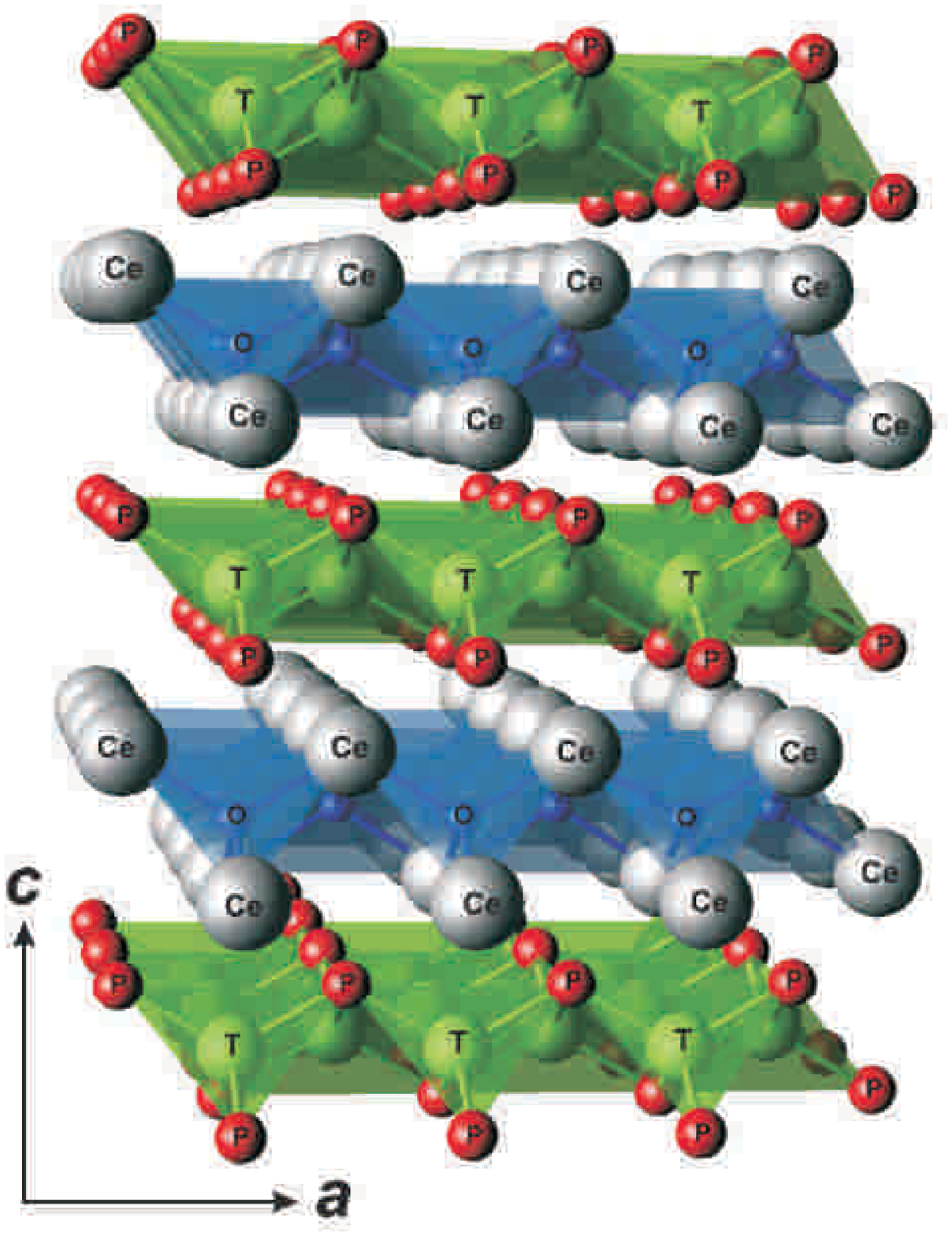}
 \caption{\label{FigStr} (Color online) Tetragonal crystal structure ($P4/nmm$) of the CeTPO compound series, showing the alternating layers of TP$_4$ and OCe$_4$ tetrahedra.}
\end{figure}

Several powder X-ray diffraction patterns recorded on a Stoe diffractometer in transmission mode using a monochromated Cu-K$_{\alpha}$ radiation ($\lambda = 1.5406$\,\AA) confirmed the formation of single phase CeTPO.  The lattice parameters refined by simple least square fitting for CeRuPO ($a=4.028(1)$\,\AA, $c=8.256(2)$\,\AA) and LaRuPO ($a=4.048(2)$\,\AA, $c=8.410(4)$\,\AA) were found to be in good agreement with the reported single crystal data. \cite{Zimmer:1995} For CeOsPO, the first refinement was performed using an internal silicon standard yielding $a=4.031(1)$\,\AA, $c=8.286(3)$\,\AA $ $, which was later confirmed in various measurements from different batches. The tetragonal structure of CeTPO with the space group \textit{P4/nmm} is shown in Fig. \ref{FigStr}, it consists of alternating layers of face sharing TP$_4$ and OCe$_4$ tetrahedra. The two dimensional nature of this lattice in which the CeO layers are well separated by TP layers, together with the presence of conduction electrons, opens up a lot of possibilities for interesting electron-mediated magnetic interactions.

\subsection{\label{CeRuPO}CeRuPO, a FM Kondo lattice system}
\begin{figure}
\includegraphics[width=10cm]{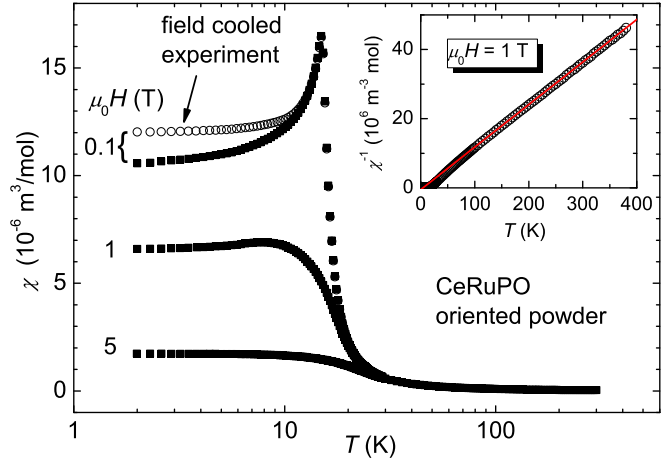}
 \caption{\label{FigMvT} (Color online) $\chi$ vs $T$ at various magnetic fields along the easy direction of oriented powder of CeRuPO. The magnetic transition into a FM ordered state is clearly visible at $T_{\rm C}=15$\,K. Open symbols indicate data from a field cooled experiment. The inset shows the inverse magnetic susceptibility together with a Curie Weiss fit ($\mu_{\rm eff}=2.3\,\mu_{\rm B}$ and $\Theta_{\rm W} = +8$\,K).}
\end{figure}

\begin{figure}
\includegraphics[width=10cm]{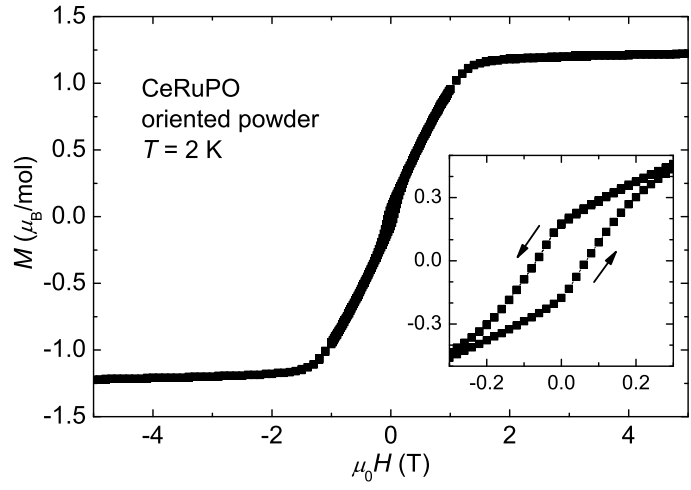}
 \caption{\label{FigMvH} Isothermal magnetization as a function of applied magnetic field at $T=2$\,K. The inset shows the small hysteresis at smaller magnetic fields.}
\end{figure}

Magnetic measurements were performed in the temperature range from $2-400$\,K in a commercial Quantum Design (QD) magnetic property measurement system (MPMS) equipped with an RSO option. The polycrystalline CeRuPO material was powdered, mixed with paraffin, and aligned in a magnetic field of $\mu_0H=5$\,T at 80$^{\circ}$C. NMR measurements discussed below suggest that this orientation corresponds to the direction $H\parallel ab$. In Fig.~\ref{FigMvT}, we show the dc-susceptibility $\chi$ as a function of temperature $T$ at three different magnetic fields. The transition into a FM ordered state is clearly visible by a sharp increase of the magnetization at $T_{\rm C}=15$\,K. The inverse susceptibility shown in the inset of Fig.~\ref{FigMvT} supports that this transition comes from local Ce$^{3+}$ moments because the data for $T>100$\,K can be fitted with a Curie Weiss law with an effective moment of $\mu_{\rm eff}= 2.3$\,$\mu_{\rm B}$/Ce close to that of the free ion moment ($2.54\,\mu_{\rm B}$/Ce). The Weiss temperature $\Theta_{\rm W}$ obtained from this linear fit is $\Theta_{\rm W} = +8$\,K. At lower temperatures $(T<50$\,K), we observe a change of the slope, resulting again in a linear behavior with the fit parameters $\mu_{\rm eff}^{{\rm low} T}= 2.2\,\mu_{\rm B}$/Ce and $\Theta_{\rm W}^{{\rm low}T} = +16$\,K, indicating FM correlations. The change of slope and of $\Theta_{\rm W}$ can be attributed to the effect of the crystal electric field (CEF). A FM state is also evidenced by the magnetization measurements at 2 K on the same sample, shown in Fig.~\ref{FigMvH}. A small but well defined hysteresis can be discerned (see inset of Fig.~\ref{FigMvH}). However, for the oriented powder the saturation magnetization $\mu_{\rm sat}= 1.2\,\mu_{\rm B}$ per Ce atom is only reached at 1.2\,T. Also the peak observed at low field in the susceptibility measurements does not correspond to the expectation for a simple ferromagnet. The height of this peak varies from sample to sample but is not solely due to domain reorientation because it was also observed in a field cooling experiment (open symbols in Fig.~\ref{FigMvT}). Presently, it is not clear whether these observations are the result of measuring powder samples of a system with a complex anisotropic behavior, like e.g., a competition between CEF and exchange interaction as in YbNiSn; \cite{Drescher:1996} or if they point to a more complex magnetic structure, e.g., a spiral with a very small propagation vector $Q$. Up to now, we could not succeed in growing single crystals large enough to measure the orientation dependence of physical properties, to determine the magnetic anisotropy of this system. However, similar behavior of the susceptibility was observed in the FM compound TbZrSb, crystallizing in a related structure. \cite{Welter:2003,Morozkin:2005}

\begin{figure}
\includegraphics[width=10cm]{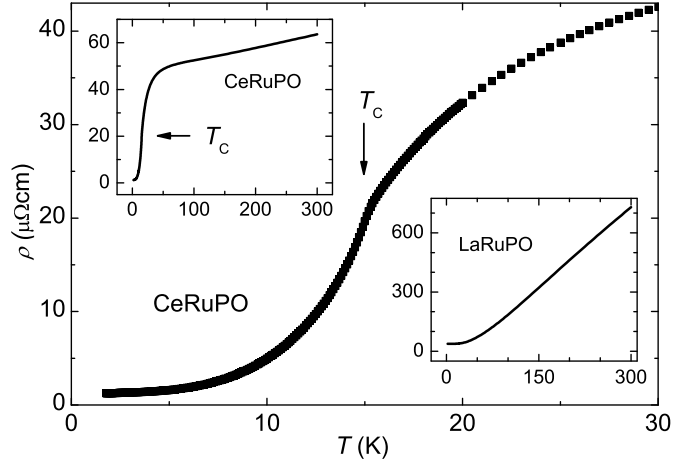}
 \caption{\label{FigRvT} Temperature dependence of the resistivity of CeRuPO. The pronounced decrease below 50\,K is attributed to coherent Kondo scattering (see upper inset). The magnetic order is visible as an additional decrease at 15\,K. In the lower inset, the temperature dependence of the resistivity of the nonmagnetic LaRuPO is shown.}
\end{figure}

AC-transport resistivity measurements were performed in a standard four-probe geometry using a commercial QD physical property measurement system (PPMS). The powdered material was pelletized and subsequently annealed at 1200$^{\circ}$C to improve crystallinity. The temperature dependence of the resistivity is plotted in Fig.~\ref{FigRvT}. Three different temperature regions can be identified: (I) Above 50\,K, the sample shows a linear behavior typical of a conventional metal. (II) Below 50\,K, there is a pronounced decrease, deviating from the linear behavior which is a distinct feature of a Kondo lattice system. This decrease can be explained by the onset of coherent Kondo scattering due to the hybridization between the localized $4f$ and the conduction electrons. (III) The FM order is clearly visible in a distinct anomaly at $T_{\rm C}=15$\,K. In the ordered state, the resistivity follows a power law dependence with $\rho\propto T^4$. All these features are reproducible, and the very good sample quality (residual resistivity $\rho_0=1.5\,\mu\Omega$cm) validates the intrinsic nature of these anomalies. In the lower inset of Fig.~\ref{FigRvT}, the resistivity of the nonmagnetic reference sample LaRuPO is shown which is typical for an usual metal without any anomalies down to 0.5\,K. Our aim to visualize the high temperature Kondo maximum in the resistivity of CeRuPO by subtracting the nonmagnetic data from the resistivity of the CeRuPO sample failed because the room temperature resistivity value of LaRuPO is with 700\,$\mu\Omega$cm one order of magnitude larger than the value of CeRuPO and makes a subtraction very arbitrary. This higher resistivity value of LaRuPO is not intrinsic, but can be ascribed to the granularity of the sample. However, the presence of Kondo interaction in CeRuPO is further supported by thermopower measurements which will be discussed below.

\begin{figure}
\includegraphics[width=10cm]{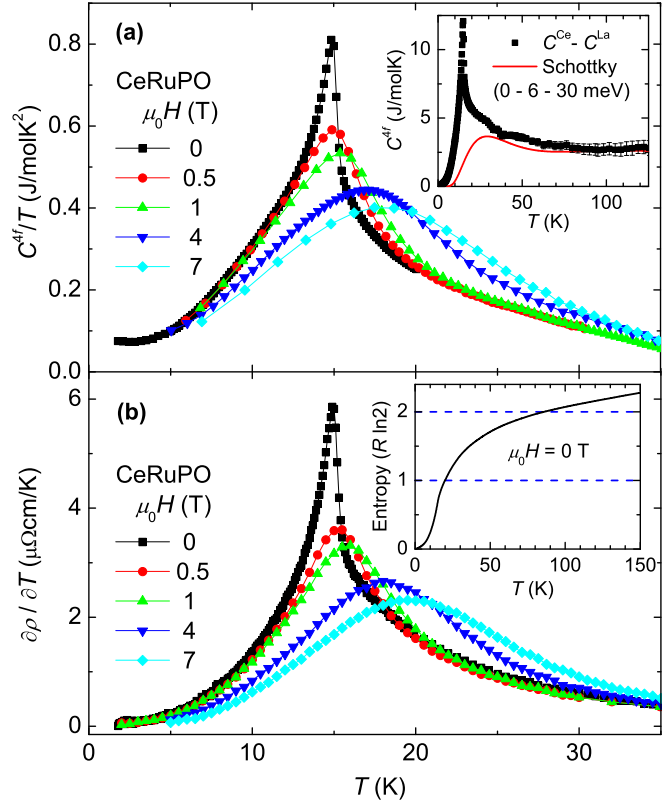}
 \caption{\label{FigCTvT} (Color online) (a) Temperature dependence of the magnetic part of the specific heat divided by temperature $C^{4f}/T$ of CeRuPO. The sharp $\lambda$-type anomaly at $T_{\rm C}$ = 15\,K corresponds to the magnetic order and shifts to higher temperatures in applied magnetic field. The inset shows $C^{4f}$ at higher temperatures, the broad Schottky peak above the transition is due to excitations of the crystal electric field levels. (b) Derivative of the resistivity of CeRuPO with respect to temperature plotted as $\partial \rho / \partial T$ vs $T$, being very similar to the specific heat data $C^{4f}/T$. The inset shows the temperature dependence of the entropy, calculated from specific heat data.}
\end{figure}

The specific heat was determined with the PPMS using a standard heat-pulse relaxation technique. The $4f$ contribution to the specific heat $C^{4f}$ of CeRuPO was obtained by subtracting the measured specific heat of the nonmagnetic reference sample LaRuPO ($C^{\rm La}$) from the specific heat of the CeRuPO polycrystals ($C^{\rm Ce}$) as $C^{4f}=C^{\rm Ce}-C^{\rm La}$. The result is shown in the upper panel of Fig.~\ref{FigCTvT}, plotted as $C^{4f}/T$ vs $T$ for different magnetic fields. A large $\lambda$-type anomaly is evident at $T_{\rm C}=15$\,K, indicating a second order transition into the magnetic state. The FM nature of the transition is supported by the field dependence of the $\lambda$-type anomaly. The transition temperature shifts to higher temperatures with increasing magnetic field; the maximum gets broader and reaches 20\,K at $\mu_0H=7$\,T. The extrapolated value of the linear part of the total specific heat $C^{\rm Ce}/T$ at $T=0$\,K gives $\gamma_0^{\rm Ce}=77$\,mJ/molK$^2$, which is enhanced by a factor of 20 compared to the value $\gamma_0^{\rm La}=3.9$\,mJ/molK$^2$ that we observe in LaRuPO. This enhancement can be attributed to the $4f$-correlation effects. The entropy gain at 20\,K is $R\ln 2$ and is slightly shifted to higher temperatures if magnetic field is applied. At higher temperatures ($T>30$\,K), $C^{4f}$ does not go to zero, indicating further excitations from the $4f$ electrons. This contribution to the specific heat is shown in the inset and can be explained by a broad Schottky anomaly, resulting from excitations to CEF levels of two excited Kramer's doublets at 6 and 30\,meV above the doublet ground state. For Ce$^{3+}$ with $J=5/2$ in a tetragonal environment, one expects for the CEF states a ground state doublet and two excited doublets. The entropy confirms low lying CEF levels because at 150\,K already 90\% of $R\ln 6$ is reached, which is the total entropy of the whole multiplet (see lower inset of Fig.~\ref{FigCTvT}). The absence of a plateau in the entropy around $R\ln 2$ indicates that the first excited CEF level is not far above $T_{\rm C}$. From the reduced jump height of the specific heat at the ordering temperature $C^{4f}-C^{\rm Schottky}\sim 9$\,J/molK, it is possible to estimate the ratio of $T_{\rm K}/T_{\rm C}\sim 0.5$. \cite{Besnus:1992} Therefore, the order of magnitude of the Kondo energy scale is $T_{\rm K}\sim 10$\,K, which is in accordance with the onset of coherent Kondo scattering at 50\,K in the resistivity. In the lower part of Fig.~\ref{FigCTvT}, we show the temperature dependence of the derivative of the resistivity, $\partial\rho/\partial T$ vs $T$. The anomaly in the resistivity shown in Fig.~\ref{FigRvT} is clearly visible in this plot, showing a sharp peak at $T_{\rm C}$. This peak behaves very similar to that in the specific heat, both at $H=0$ and $H>0$. Such a pronounced similarity has been reported and analyzed by Campoy \textit{et al.} \cite{Campoy:2006}

\begin{figure}
\includegraphics[width=10cm]{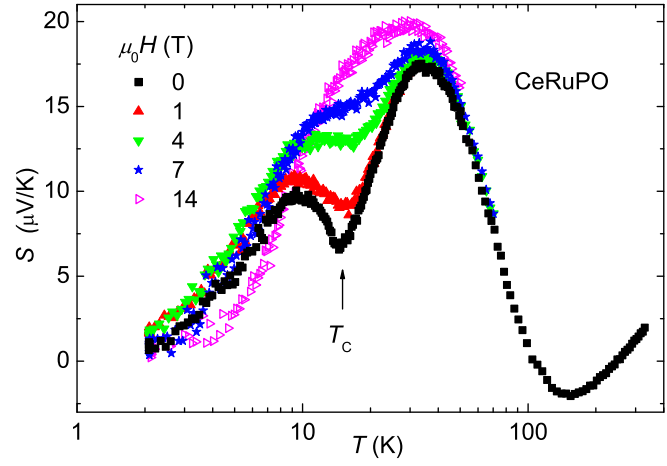}
 \caption{\label{FigTTO} (Color online) Thermoelectric power vs temperature of CeRuPO for different magnetic fields.}
\end{figure}

To further study the influence of the Kondo interaction on the transport properties of CeRuPO, we performed thermal transport (TT) measurements on bar shaped polycrystalline samples in a four-probe geometry, using the TT option of the PPMS. Among other transport measurements, the thermopower is of particular interest because it is extremely sensitive to any variation of the density of states at the Fermi level and exhibits large values (10-100\,$\mu$V/K) for Kondo lattice systems. \cite{Garde:1995} The temperature dependence of the thermopower $S(T)$ of CeRuPO is shown in Fig.~\ref{FigTTO}, plotted on a logarithmic temperature scale. $S(T)$ has a broad minimum at 150\,K (which is negative), followed by a strong increase below 100\,K to a maximum at 35\,K with absolute values as high as 18\,$\mu$V/K. A pronounced minimum at $T_{\rm C}$ separate this first maximum from a second one at 10\,K. At lower temperatures ($T<10$\,K), the thermopower goes within the error bars linearly to zero. This behavior of $S(T)$ was reproduced in a different sample. Large positive maxima are common in many Ce-based Kondo lattice systems and can be ascribed to the Kondo scattering of the different CEF levels. \cite{Amato:1989, Zlatic:2005} Since the CEF maximum in the thermopower is usually observed around half of the splitting energy, \cite{Wilhelm:2004} the peak at 35\,K in CeRuPO can therefore be nicely attributed to the first excited CEF level at 6\,meV, visible also in the magnetic part of the specific heat. The peak position is unaffected by a magnetic field, which confirms the CEF origin. At lower temperatures ($T<30$\,K), the $T$-dependence of $S(T)$ results from an interplay between Kondo effect and FM ordering. Because of the lack of FM Kondo systems, the effect of this interplay on $S(T)$ has not yet been studied. The only FM Kondo lattice system where thermopower measurements were performed is the related compound CeAgSb$_2$. \cite{Houshiar:1995} However, CeAgSb$_2$ presents a strong decrease of $S(T)$ below $T_C$, in contrast to the increase observed in CeRuPO. The shift of the minimum in $S(T)$ of CeRuPO to higher temperatures with increasing magnetic field indicate that it arises from FM correlations. A likely explanation of the temperature dependence of $S(T)$ is that in the absence of FM correlations, $S(T)$ would present a broad maximum due to both Kondo effect and CEF splitting, \cite{Zlatic:2003, Hartmann:2006} as the involved energy scales ($T_{\rm K}\sim T_{\rm CEF}$) are close to each other. FM inter-site fluctuations lead to a reduction of $S(T)$ at $T_{\rm C}$, this reduction becoming weaker and shifting to higher $T$ with increasing magnetic field. Then, the field dependence of the low temperature maximum would just be a consequence of the field dependence of the minimum. To understand the observed behavior in more detail, theoretical predictions for $S(T,H)$ in FM Kondo systems are necessary. 

\begin{figure}
\includegraphics[width=10cm]{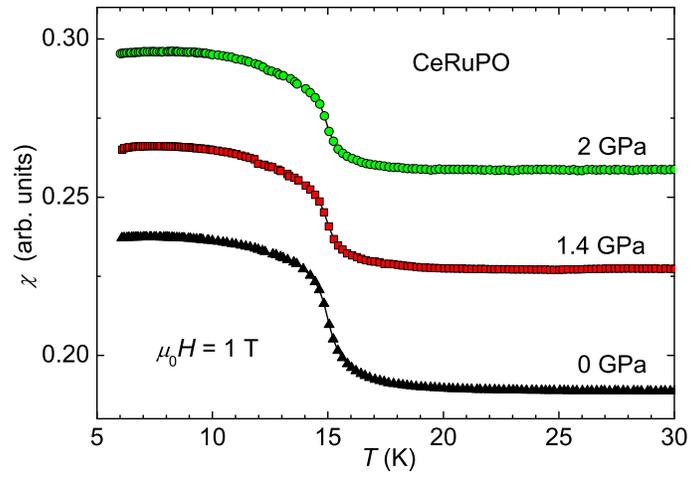}
 \caption{\label{FigPress} (Color online) $\chi$ vs $T$ measured in a magnetic field of 1\,T at atmospheric pressure and under an applied pressure of $p=1.4$\,GPa and 2\,GPa, respectively. The data for the different pressures are shifted with respect to each other for clarity.}
\end{figure}

The dc-susceptibility $\chi(T)$ under pressure up to 2\,GPa was measured in a QD-MPMS using a diamond (0.25 carat) anvil cell with a culet size of 1.5\,mm. As gasket we used a CuBe disk with a bore of 800\,$\mu$m in diameter. Daphne oil served as pressure transmitting medium. The pressure inside the cell was determined by the inductively measured shift of the superconducting transition temperature of lead (Pb) placed next to the CeRuPO sample, relative to a Pb sample outside of the cell. In Fig.~\ref{FigPress}, $\chi(T)$ is shown as function of $T$ for different pressures. For 1.4\,GPa and 2.0\,GPa no significant changes in the temperature dependence of $\chi(T)$ are observed; and therefore, $T_{\rm C}$ stays constant up to 2.0\,GPa. In Ce based Kondo lattice systems, pressure is expected to suppress magnetic order because the volume of the nonmagnetic Ce$^{4+}$ ion is smaller. However, it was observed that in a FM Kondo lattice system there can be a broad pressure range where $T_{\rm C}$ only slightly changes. \cite{Larrea:2005} Further experiments with an extended pressure range are needed to study the influence of pressure in CeRuPO in more detail.

\subsection{\label{CeOsPO}CeOsPO, a AFM Kondo lattice system}
\begin{figure}
\includegraphics[width=10cm]{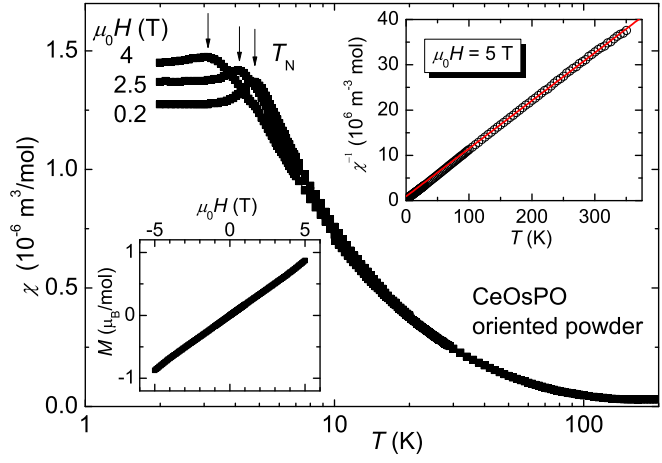}
 \caption{\label{FigMvTOs} (Color online) $\chi$ vs $T$ for three different magnetic fields along the easy direction of oriented powder of CeOsPO. The magnetic transition into a AFM ordered state is visible at $T_{\rm N}=4.5$\,K. The upper inset shows the inverse magnetic susceptibility together with a Curie Weiss fit ($\mu_{\rm eff}= 2.45\,\mu_{\rm B}$ and $\Theta_{\rm W} = -9$\,K). In the lower inset, the magnetization is plotted vs magnetic field at $T=2$\,K. No metamagnetic transition is visible up to 5\,T.}
\end{figure}

Another way to suppress magnetic order is to dope the system chemically with an isoelectronic element which leads to stronger hybridization and favors the Kondo interaction. Since $5d$ metals usually cause stronger hybridization than $4d$ metals, we have synthesized CeOsPO. As discussed above, the volume of the unit cell changes only slightly and is 0.5\% larger than the unit cell of CeRuPO. The inverse susceptibility of CeOsPO is shown in the upper inset of Fig.~\ref{FigMvTOs}. The Curie Weiss fit with $\mu_{\rm eff}= 2.45\,\mu_{\rm B}$/Ce close to that of the free ion moment for $T>100$\,K again shows that Ce is in the trivalent state. The Weiss temperature $\Theta_{\rm W} = -9$\,K obtained from this linear fit indicates AFM correlations. The temperature dependence of $\chi$ (Fig.~\ref{FigMvTOs}) is typical for an AFM ordered system because the magnitude of the susceptibility is nearly field independent. Further on, a peak marks the onset of AFM order at $T_{\rm N}=4.4$\,K for $\mu_0H=0.2$\,T. For higher magnetic fields, this ordering temperature decreases; for clarity, we only show the data points for three selected magnetic fields. At $T=2$\,K the magnetization increases linearly with the magnetic field, no metamagnetic transition is visible up to $5$\,T (see lower inset of Fig.~\ref{FigMvTOs}).

\begin{figure}
\includegraphics[width=10cm]{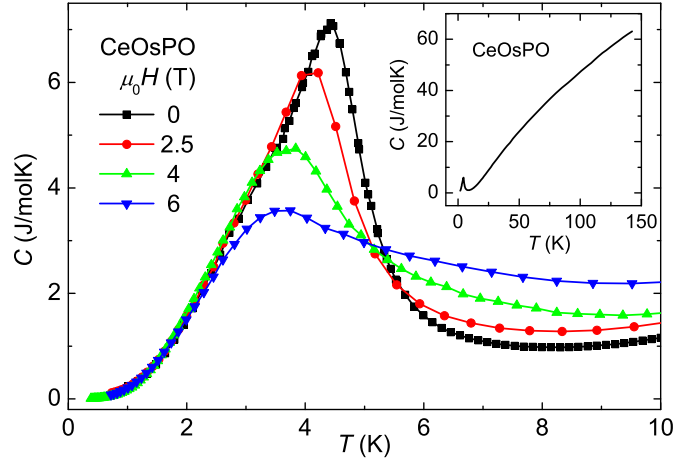}
 \caption{\label{FigCvTOs} (Color online) Temperature dependence of the specific heat $C$ of CeOsPO. The phonon contribution is not subtracted in this representation. The large anomaly at $T_{\rm N}$ = 4.4\,K corresponds to the magnetic order and shifts to lower temperatures in applied magnetic field. The inset shows the specific heat of CeOsPO in an extended $T$ range.}
\end{figure}

In Fig.~\ref{FigCvTOs}, we present the specific heat $C$ of CeOsPO, plotted as function of temperature. We did not subtract any phonon contribution which at low temperatures is negligible compared to the large mean field type anomaly at $T_{\rm N}=4.4$\,K, indicating that the magnetic order is a bulk property of CeOsPO. The entropy gain involved in the magnetic transition is close to $R\ln 2$ at 10\,K. The transition becomes broader if magnetic field is applied and $T_{\rm N}$ shifts to lower temperatures, in accordance with the AFM nature of the transition. No other transition has been found up to 150\,K as can be seen in the inset of Fig.~\ref{FigCvTOs} which shows $C$ in the whole measured temperature range. Up to now, we were not able to perform any reproducible transport measurements of CeOsPO due to difficulties in preparing a compact sample of sufficient size.

\subsection{\label{NMR} $^{31}$P NMR on CeTPO}
\begin{figure}
\includegraphics[width=10cm]{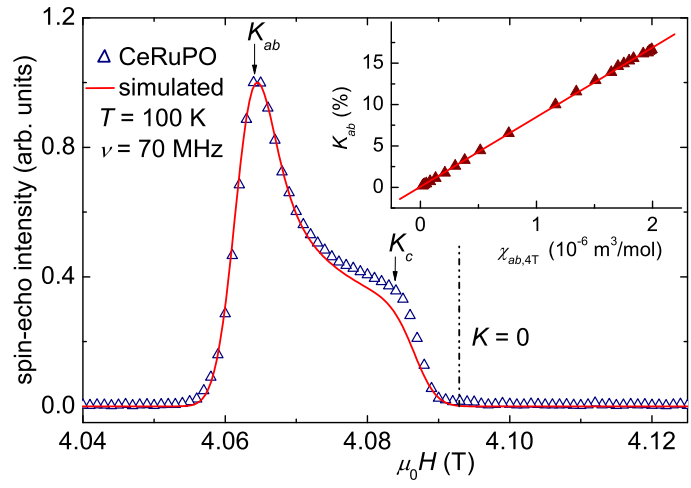}
 \caption{\label{FigNMR1} (Color online) Spectrum of CeRuPO at  $T = 100$\,K and $\nu$ = 70\,MHz. The solid line describes a simulated spectrum. The inset shows the linear behavior of $K_{ab}$ vs $\chi_{ab}$.}
\end{figure}

We confirmed the conclusions drawn from thermodynamic and transport measurements, using results obtained with a microscopic technique, namely  $^{31}$P nuclear magnetic resonance (NMR). These measurements were performed on polycrystalline samples with a standard pulsed NMR spectrometer at a fixed frequency of about 70\,MHz and in the temperature range between 2 and 300\,K. The NMR field sweep spectra were obtained by tracing the intensity of the echo as a function of field, using a common spin-echo sequence. NMR shift values were determined by using H$_3$PO$_4$ as a reference ($K = 0$, line in Fig.~\ref{FigNMR1}). The measurement of the spin-lattice relaxation time was carried out with a saturation recovery sequence. The $^{31}$P spectrum ($I = \frac{1}{2},\, \gamma = 17.1027$\,MHz/T) in Fig.~\ref{FigNMR1} shows a typical NMR powder pattern expected for axial symmetry (tetragonal in this case). The simulation of the spectra at a fixed temperature allows the determination of the anisotropic Knight shift components $^{31}K_{ab}(T)$ and $^{31}K_{c}(T)$, corresponding to the $H \perp c$ and $H \parallel c$ directions in the crystal. The assignment of these components is clearly given by the fact that the highest intensity in the powder pattern is attributed to the $H \perp c$ direction. A plot $K_{ab}$ vs $\chi_{ab}$ (measured on oriented powder in a magnetic field of 4\,T), see inset of Fig.~\ref{FigNMR1}, evidences a linear dependence, as expected from the relation $K_{ab}=(A_{\rm hf}/\mu_{\rm B} N_{\rm A})\chi_{ab}$. \cite{Carter:1977} From the slope we determined the hyperfine coupling constant $^{31}A_{\rm hf}$ for CeRuPO, $^{31}A_{\rm hf} =6$\,kOe/$\mu_{\rm B}$; a value comparable to $^{31}A_{\rm hf}=8.73$\,kOe/$\mu_{\rm B}$ reported for CeP another P containing Kondo system. \cite{Kobayashi:1996} $^{31}K_{ab}(T)$ is positive and reaches its maximum value of about +17\% at 2\,K. When the FM order sets in ($T_{\rm C,4\,T}\sim 18$\,K), the line broadens, but the mean value of the shift stays constant with decreasing temperature which is expected for a ferromagnet. In contrast to CeRuPO, $^{31}K_{ab}$ for the Os compound is negative, and a hyperfine coupling constant of $^{31}A_{\rm hf} = -2.4$\,kOe/$\mu_{\rm B}$ could be determined. \cite{Bruening}

\begin{figure}
\includegraphics[width=10cm]{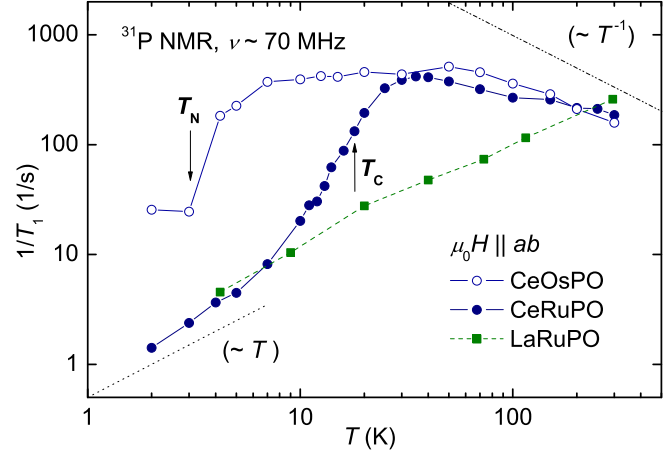}
 \caption{\label{FigNMR2} (Color online) Spin lattice relaxation rate $^{31}(1/T_1)$ of CeRuPO (filled circles) and CeOsPO (open circles), together with $^{31}(1/T_1)$ of the nonmagnetic LaRuPO (filled squares).}
\end{figure}

Spin-lattice relaxation measurements were carried out at the $^{31}K_{ab}(T)$ position in the spectra (see Fig.~\ref{FigNMR1}). The spin relaxation rate $T_1^{-1}(T)$ shown in Fig.~\ref{FigNMR2} is quite similar in the two compounds at high temperatures, while below 40\,K, it differs significantly in accordance to the different behavior observed in macroscopic measurements. At 300\,K, $T_1$ has almost the same value in both systems. With decreasing temperature, $T_1^{-1}$ first increases slightly with a smaller slope in CeRuPO than in CeOsPO where it approximately follows a power law $T_1^{-1}\propto T^{-1}$. Such a behavior is typical for well localized $4f$ systems for $T>>T_{\rm K}$. \cite{Nakamura:1996, Buttgen:1996} In CeRuPO, $T_1^{-1}(T)$ presents a well defined maximum around 35\,K before decreasing significantly at lower temperatures and merging in a linear in $T$ behavior below 10\,K. Since the temperature of the maximum corresponds to the onset of coherent Kondo scattering in $\rho(T)$, this maximum could be attributed to the onset of the Kondo effect. However, the observation that this maximum shifts to lower temperature and becomes more pronounced with decreasing magnetic field indicates that it is also related to the onset of FM correlations. \cite{Bruening} A separation of both effects is presently not possible. In contrast, in CeOsPO $T_1^{-1}$ becomes $T$-independent between 60 and 6\,K before it starts to decrease towards lower temperatures. Such a plateau like behavior is expected in Kondo lattice systems in the temperature range just above $T_{\rm K}$. \cite{Nakamura:1996, Buttgen:1996} This would suggest a Kondo scale of the order of 5\,K in CeOsPO. The decrease of $T_1^{-1}$ in CeOsPO below 6\,K ends in a pronounced drop at the temperature corresponding to $T_{\rm N}$ at the field strength of the NMR experiment. The absence of a critical fluctuation induced maximum in $T_1^{-1}$ at $T_{\rm C}$ in CeRuPO and at $T_{\rm N}$ in CeOsPO might be due to the broadening effect of the applied field in the former one and/or to the cancellation of these fluctuations at the P-site in the latter one. We included in Fig.~\ref{FigNMR2} the temperature dependence of $T_1^{-1}$ of the nonmagnetic reference compound LaRuPO. A common approach in the analysis of $T_1$ in rare earth transition metal compounds is to separate the contributions to the spin relaxation, in a contribution of the local moment fluctuations which is related to the bulk susceptibility, and a contribution of the itinerant conduction electrons which follows a Korringa behavior $T_1^{-1}\propto T$. This approach seems to be quite appropriate for the present compounds, since the difference between $T_1^{-1}(T)$ of CeRuPO and LaRuPO first increases strongly with decreasing temperature passes through a maximum around 30\,K before vanishing at low $T$. A more precise analysis of the NMR results shall be performed in a paper devoted to NMR experiments. \cite{Bruening}

\section{\label{Theory}Band structure calculations}
To gain deeper insight into the electronic structure of CeTPO, band structure calculations were performed using the full-potential local-orbital minimum basis code FPLO (version 5.00-19) \cite{FPLO} within the local (spin) density approximation (L(S)DA). In the scalar relativistic calculations the exchange and correlation potential of Perdew and Wang \cite{PW} was employed. As basis set, Ce ($4p4d4f5s5p$ / $6s6p5d$), Ru ($4s4p$ / $5s5p4d$), Os ($4p4d4f5s5p$ / $6s6p5d$), P ($2s2p$ / $3s3p3d$), and O ($2s2p3d$) where chosen as semi-core/valence states. All lower lying states were treated fully relativistic as core states. The inclusion of semi-core states was forced by their non-negligible overlap due to the large extension of their wave functions. The formally unoccupied P and O 3$d$ states were included to improve the completeness of the basis set. The extension of the valence orbitals is controlled by an additional confining potential ($r/r_0$) and gets optimized with respect to the total energy. \cite{HE} To treat the strong Coulomb correlation of Ce $4f$ electrons in mean field approximation, LSDA+$U$ (in the around mean field double counting scheme) was applied with $U=6.1$\,eV and $J=0.7$\,eV ($F_2=8.34$\,eV, $F_4=5.57$\,eV and $F_6=4.12$\,eV). \cite{SP} In the ferromagnetic cell 546 $k$-points, and in the antiferromagnetic $\sqrt2 \cdot \sqrt2$ super cells 640 $k$-points in the irreducible part of the Brillouin zone (BZ), respectively, were used as converged $k$-mesh. These $k$-mesh sizes ensures a numerical accuracy of the total energy at least one order of magnitude better than change in energy for the investigated physical quantities. Because of the very small energy scales between different magnetic states, CeRuPO was also calculated with the LAPW code WIEN2k \cite{WIEN} for comparison with an alternative full potential scheme. \cite{Rosner} The resulting electronic densities of states, band structures, and total energy differences were basically identical for these two band structure codes.

\begin{figure}
\includegraphics[width=8.5cm]{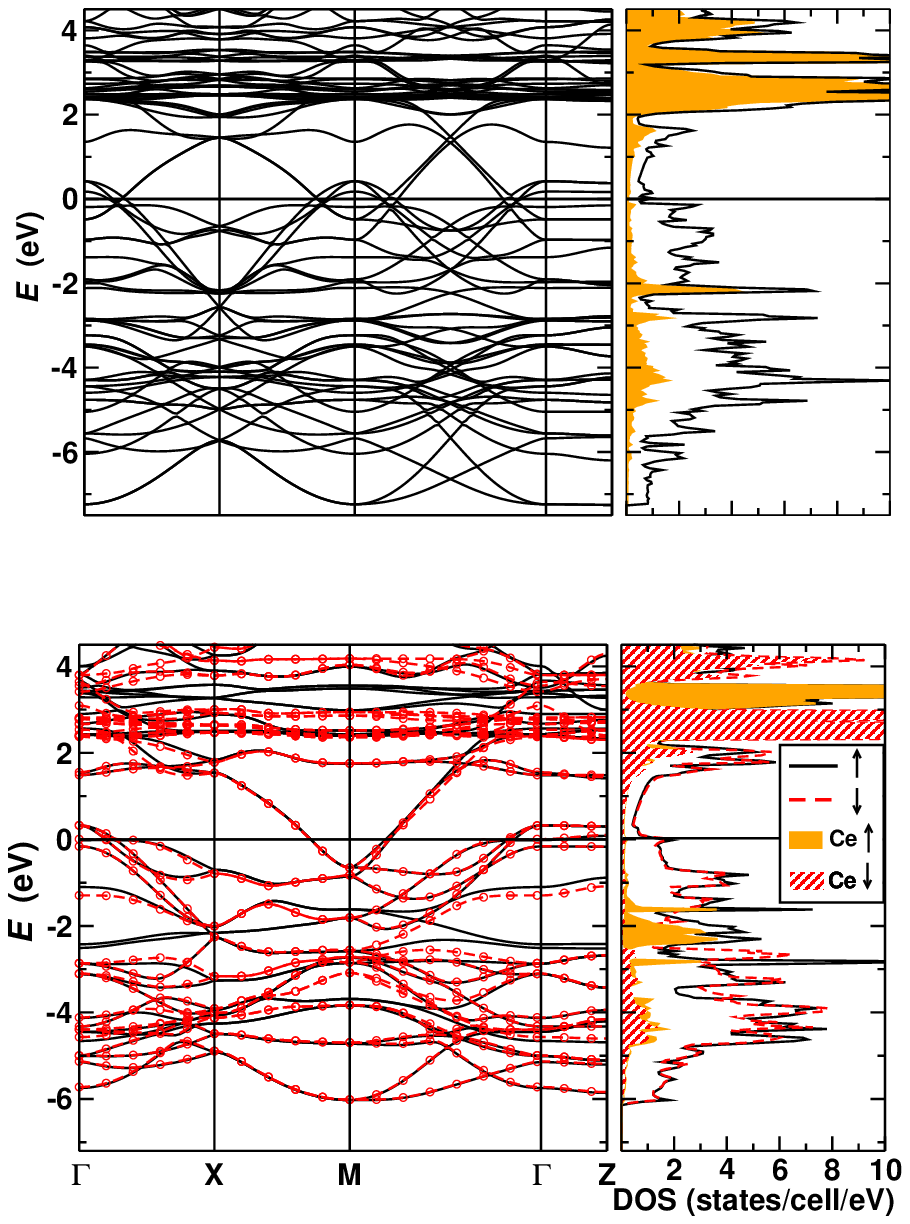}
\caption{(Color online) Band structure and total as well as partial Ce densities of states for the FM state of CeRuPO (lower panel) and the G-AFM state of CeOsPO (upper panel). The symmetry points of the BZ correspond to the primitive cell for better comparability. The Ce contributions to the densities of states are indicated by different fillings. The spin down bands for CeRuPO are indicated by dashed (red) lines with small circles.}
\label{DOSEN}
\end{figure}

For both the Ru and the Os compound, we calculated the total energies and the electronic structures for three different magnetic states: (I) fully ferromagnetic arranged Ce atoms (FM), (II) ferromagnetic arranged Ce atoms in the plane but antiferromagnetic between the planes (A-AFM), and (III) antiferromagnetic arrangement of the Ce atoms in the plane and between the planes (G-AFM). For CeRuPO, we found the FM state to be lowest in energy, followed by the G-AFM ($+94.6$\,meV) and the A-AFM state ($+94.9$\,meV). In contrast, for CeOsPO the G-AFM state was slightly favored in energy, followed by the A-AFM ($+0.08$\,meV) and the FM state ($+1.2$\,meV). This result is in nice agreement with the experimental observation although the rather small energy differences for the Os compound should be taken with care, especially for the difference between the two AFM order types. The corresponding densities of states (DOS) and band structures (FM for CeRuPO and G-AFM for CeOsPO) are shown in Fig.\,\ref{DOSEN}.

\begin{figure}
\includegraphics[width=8.5cm,angle=-90]{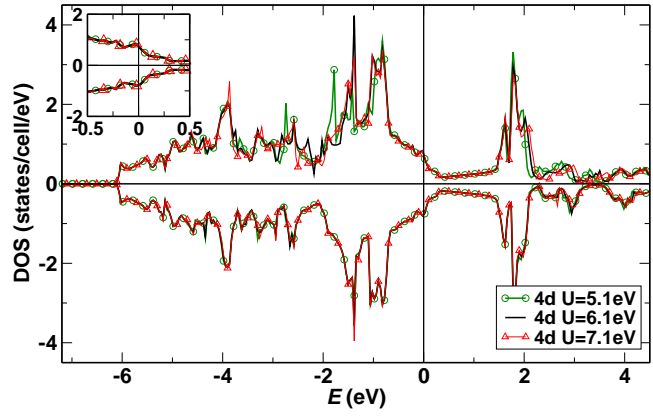}
\caption{(Color online) Partial Ru $4d$ DOS of CeRuPO for three different choices of the Coulomb repulsion parameter $U$. The inset shows a narrow region around the Fermi level where the curves are identical within the line widths. Small differences only occur in energy regions with significant Ce $4f$ contributions (see Fig.\ \ref{DOSEN}).}
\label{LDU}
\end{figure}

\begin{figure}
\includegraphics[width=8.5cm]{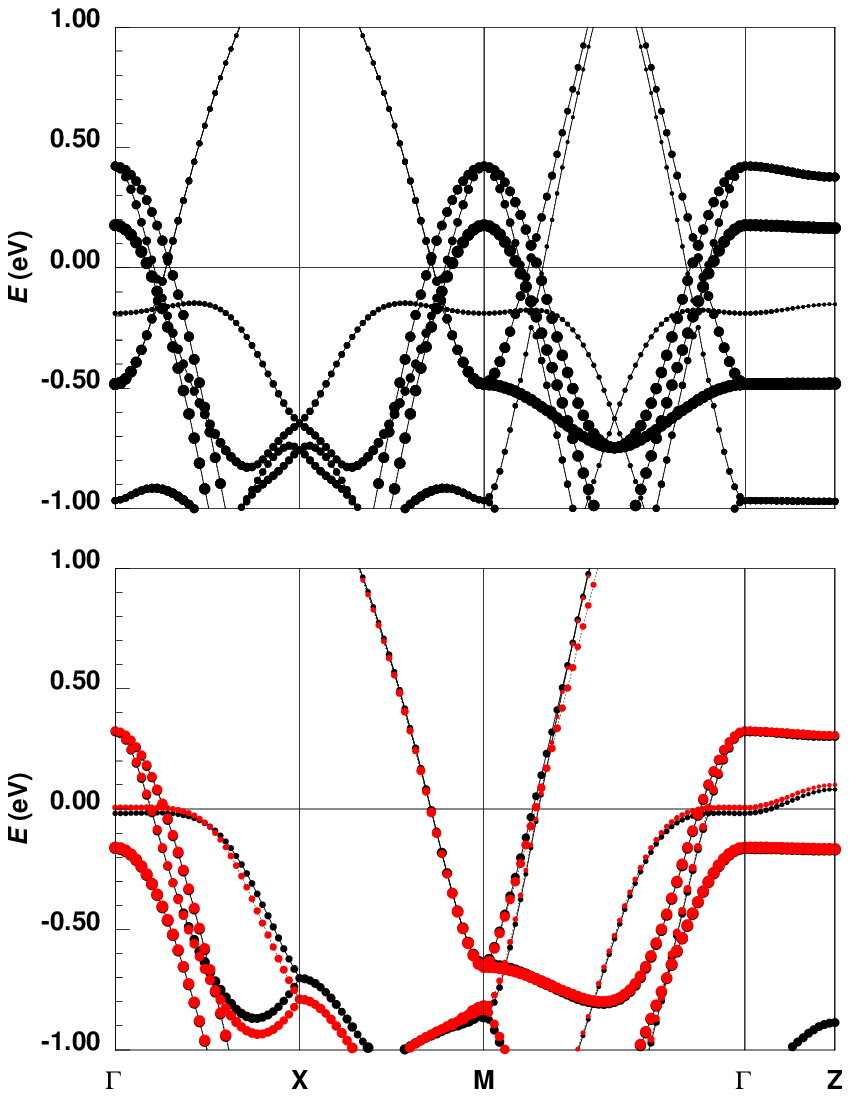}
\caption{(Color online) Band structure close to the Fermi level for CeRuPO (lower panel) and CeOsPO (upper panel). The symmetry points of the BZ correspond to the original non-magnetic cell.  The transition metal $d$ band characters are indicated by the size of the filled circles. The two spin
directions for Ru are shown by black and grey (red) symbols.}
\label{BANDS}
\end{figure}

At first glance both band structures and DOS's look very similar. Note that for the Os compound (see Fig.\,\ref{DOSEN}, upper panel) the number of bands is doubled due to the doubled number of atoms in the magnetic super cell, resulting in a folding of the band structure ($\Gamma$ and M are identical in the chosen notation of symmetry points referring to the original non-magnetic unit cell). The filled valence states originate essentially from P and O $p$ and from the transition metal states. The states between about -2\,eV and 0.5\,eV are dominated by the transition metal $d$ electrons. In the LDA calculations (not shown), the strong Coulomb repulsion of the Ce $4f$ electrons is largely underestimated, resulting in an unrealistic peak of the partial Ce $4f$ DOS at the Fermi level. The application of LSDA+$U$ splits the Ce $4f$ states and yields the (qualitatively) correct physical picture of a Ce$^{3+}$ state in both compounds where the occupied and the unoccupied Ce $4f$ states show a split of the order of $U$. Thus, the magnetic Ce spin moment is about $1\,\mu_B$ for both compounds. All other atoms show negligible spin polarization, including Ru and Os. This can be seen from the very small split of the Ru related bands between -2\,eV and 0.5\,eV (see Fig.\,\ref{DOSEN}, lower panel), especially close to the Fermi level. This picture is basically independent from the actual choice of the parameter $U$ (in a reasonable range). Fig.\,\ref{LDU} shows the partial $4d$ DOS for Ru for three representative $U$ values (6.1\,eV\,$\pm 1$\,eV). \cite{SP} The dependence on the chosen $U$ is negligible in the region around the Fermi level relevant for the low lying excitations, especially the metallic properties. This justifies the application of the LSDA+$U$ approximation although the actual value of $U$ may vary slightly. Therefore, in a first approximation the magnetic interaction in the two compounds can be understood from an RKKY type mechanism with rather localized Ce $4f$ states coupled by the itinerant transition metal $d$ electrons.  On the other hand one should keep in mind that all dynamic correlation effects are neglected in the LSDA+$U$ approximation and would require a more sophisticated theoretical treatment. Fig.\,\ref{BANDS} shows the detailed band structure close to the Fermi level for CeRuPO (lower panel) and CeOsPO (upper panel) together with the transition metal $d$ band characters. Note again the doubled number of bands for the Os compound due to the magnetic super cell, otherwise the resulting bands are very similar. Neglecting the small spin split for the Ru $4d$ related bands in CeRuPO, we observe a very close similarity with the related superconductor LaFePO. \cite{Lebegue:2007}

The characteristic feature in the CeRuPO and CeOsPO band structures is the strongly pronounced two dimensionality. The in-plane dispersion (along $\Gamma$-X-M-$\Gamma$) exceeds the out of plane dispersion (along $\Gamma$-Z) by far more than an order of magnitude. The underlying reason is the layer type crystal structure, resulting in a rather weak hybridization between adjacent metallic T-P layers. The interaction along the $z$ direction is mainly mediated by the O $2p$ orbitals of the otherwise non-metallic Ce-O layers. On the other hand, the slightly different oxygen hybridization in both compounds is most likely the reason for the different magnetic ground states. Whereas in CeOsPO the bands crossing the Fermi level are mostly dispersionless along $\Gamma$-Z, one of the bands in CeRuPO shows sizable dispersion along the z direction (see Fig.\,\ref{BANDS}). The corresponding band in the Os compound is slightly lower in energy and therefore less relevant for the magnetic interaction. Compared to the other T-derived bands, the T-$d$ character of this band is somewhat reduced due to the hybridization with P and O $p$ electrons.  Due to the two dimensional electronic structure, the resulting Fermi surfaces (not shown) are almost cylindrical tubes in the case of CeOsPO. In CeRuPO one of the cylinders closes due to the dispersion along the $z$-direction and forms an ellipsoid around the Z point like in LaFePO. \cite{Lebegue:2007} To understand the magnetic properties of the two compounds in more detail, further more sophisticated studies, including the investigation of the role of nesting properties, are necessary.

\section{\label{Discuss}Discussion}
Our results on CeRuPO and CeOsPO evidence three interesting features which shall now be discussed: The FM ordering and the Kondo behavior in CeRuPO, and the quasi two dimensionality. In the absence of more detailed information because of lack of single crystals or larger samples for neutron scattering experiments, a deeper insight into the magnetic exchange interaction in these systems can be gained by a comparison with RTX compounds (R\,=\,rare earth, X\,=\,Si,\,Ge) crystallizing in the CeFeSi structure. The arrangement of the R, T, and X atoms in this structure is identical to the presented compounds, the only difference is the insertion of O atoms between adjacent Ce-planes. B. Chevalier and B. Malaman \cite{Chevalier:2004} discussed the magnetic order in these RTX compounds using a simple model with three exchange interactions:  $J_0$ within an R-plane (this corresponds to the next nearest R-R-neighbor), $J_1$ between R atoms separated by a TX-layer, and $J_2$ between adjacent R-planes sandwiching an O-layer (this corresponds to the nearest R-R-neighbor). Systematic studies of these RTX compounds showed that $J_0$ is always positive (FM), while $J_1$ and $J_2$ change sign depending on the T element. $J_1$ and $J_2$ are positive in the RFeSi light rare earth compounds, while $J_1<0$, $J_2>0$ for T\,=\,Co,\,Ti, and $J_1>0$, $J_2<0$ for T\,=\,Ru. The only exception with both $J_1$ and $J_2$ negative is CeCoGe. Thus, for this type of arrangement of R, T and X atoms there is a clear predominance of FM exchange. Notabene strong ferromagnetism, with $T_{\rm C} = 374$\,K in GdTiGe, has also been observed in the related CeScSi structure which differs only in the long range stacking of the layers. \cite{Skorek:2001} Therefore, the observation of a FM state in CeRuPO is not surprising and can be related to an atomic arrangement promoting FM interactions. Comparing the FM state in CeRuPO with the AFM structure found in RRuSi(Ge), the inclusion of an O atom between adjacent Ce-layers switches $J_2$ from AFM to FM. In absence of more detailed information, it is yet not clear whether the AFM structure in CeOsPO is due to a negative $J_1$ or a negative $J_2$. A preliminary analysis of the change in the anisotropy of the hyperfine coupling constant between CeRuPO and CeOsPO suggest that at least $J_1$ is becoming AFM in the latter one since P is along the corresponding exchange path. \cite{Bruening}

As already stated, only very few FM Kondo lattices have been reported previously. None of them show such a strong decrease in the resistivity starting far above $T_{\rm C}$ as we observed in CeRuPO. As an example YbNiSn, CeAgSb$_2$, \cite{Sidorov:2003} and CeSi$_{1.71}$ \cite{Pierre:1990} present a weak increase in $\rho(T)$ below $50-100$\,K ending in a maximum at a temperature $T_{\rm max}$ which is at or only slightly above $T_{\rm C}$. Thus $\rho(T_{\rm C})$ is only slightly smaller than $\rho(T_{\rm max})$. In contrast, $\rho(T)$ in CeRuPO has at $T_{\rm C}$ decreased to less than 50\% of the value at 50\,K which would roughly correspond to the position of the maximum after subtraction of the phonon scattering. Thus spin flip scattering is already significantly suppressed at $T_{\rm C}$, indicating that the Kondo fluctuations at $T_{\rm C}$ are likely much stronger than in the previously known FM Kondo lattices. The most appropriate compound for a detailed comparison is the closely related CeAgSb$_2$ \cite{Sidorov:2003} which crystallizes in the same structure type as CeTPO. However, the much larger size of Sb compared to O and P leads to quite different Ce-Ce, Ce-T, and Ce-P distances. Nevertheless, CeAgSb$_2$ also orders ferromagnetically at $T_{\rm C}$ = 9.6 K. While at ambient pressure $\rho(T)$ in CeAgSb$_2$ decreases by only a few percent between $T_{\rm max}$ and $T_{\rm C}$ under an isostatic pressure larger than 2.4\,GPa, its temperature dependence becomes comparable to that in CeRuPO at $p = 0$; although, the onset of coherence stays at a much smaller temperature, indicating a smaller $T_{\rm K}$. The ratio RR$_{T_{\rm C}}$ = $\rho(T_{\rm max})/\rho(T_{\rm C})= 2.2$ in CeAgSb$_2$ at $p = 3.26$\,GPa which is the highest pressure where the FM transition is still clearly visible is close to RR$_{T_{\rm C}}=2.3$ in CeRuPO at ambient pressure, suggesting a comparable $T_{\rm K}/T_{\rm C}$ ratio under these conditions. However, because of the much lower $T_{\rm C}=2.4$\,K at $p=3.26$\,GPa in CeAgSb$_2$ this also corresponds to a much lower absolute value of $T_{\rm K}$. A smaller $T_{\rm K}$ in CeAgSb$_2$ is further evidenced by the weaker thermopower. \cite{Houshiar:1995} The larger $T_{\rm K}$ in CeRuPO is in accordance with general rules for the strength of hybridization between $f$ and conduction electrons first proposed by D.D. Koelling \textit{et al.} \cite{Koelling:1985} Both the smaller size of P compared to that of isoelectronic Sb, and the shift of the $d$-states closer to the Fermi level upon exchanging Ag by Ru are expected to lead to a stronger hybridization. One can also suspect that the replacement of one Sb by O has the tendency to push Ce towards a higher oxidation state. The effect of pressure on CeRuPO and the way in which a further increase of $T_{\rm K}$ suppress the FM order shall be an important issue for future experiments.

Quasi two dimensional (2D) intermetallic Kondo lattices have attracted considerable attention in the past years. The main reason was the discovery of a rather high superconducting transition temperature $T_{\rm c}=2.3$\,K in CeCoIn$_5$; the layered structure of this compound is suspected to be at the origin of the enhanced $T_{\rm c}$. \cite{Petrovic:2001} However, the 2D character of these 115 compounds is not very pronounced; the slope of the upper critical field at $T_{\rm c}$ e.g. shows only a rather small anisotropy of a factor of two. Only one of the three sheets of the Fermi surface of CeCoIn$_5$ has a clear 2D character while the two others are clearly 3D. \cite{Settai:2001} Thus, intermetallic heavy Fermion compounds with a better defined 2D character are of strong interest. The results of our LDA+$U$ calculations as well as those on LaFePO \cite{Lebegue:2007} suggest that the electronic states at the Fermi level of the presented RTPO have a stronger 2D character than in the CeTIn$_5$ compounds, since at least four of the five sheets of the Fermi surface are 2D. Thus the growth of single crystals large enough for allowing the study of the anisotropy of the transport and electronic properties shall be another important issue for future research.

\section{\label{Concl}Conclusions}
In summary, we have shown that the CeTPO (T\,=\,Ru,\,Os) compound series presents an interesting class of Ce-based Kondo lattice systems at the border between intermetallic and oxide materials with a layered structure and strong tendency to ferromagnetism. CeRuPO seems to be a nice example of a FM Kondo lattice ($T_{\rm C}=15$\,K, $T_{\rm K}\sim10$\,K). Together with the antiferromagnetic CeOsPO ($T_{\rm N}=4.4$\,K), these two compounds are ideally suited to study the difference between FM and AFM Kondo lattices. This change from the FM state in CeRuPO to the AFM state in CeOsPO is induced by only minor changes in the crystallographic parameters. However, further pressure experiments on CeRuPO are necessary to investigate the interplay between Kondo interaction and development of long range magnetic order. The yet unique combination of quasi two dimensional electronic properties and strong FM correlations could lead to new kind of quantum critical behavior. Furthermore, a systematic study of related derivatives by replacing the T atoms with other $d$ elements brings up a lot of possibilities for further interesting materials.

\section*{Acknowledgements}
The authors thank U. Burkhardt, P. Scheppan and G. Auffermann for chemical analysis of the samples, N. Caroca-Canales and R. Weise for technical assistance in sample preparation, U. K\"ohler for support concerning the thermopower measurements, and C. Klausnitzer for help in preparing the pressure cell. The Deutsche Forschungsgemeinschaft (Emmy Noether Programm, SFB 463) is acknowledged for financial support.

$ $
\newpage

$ $
\newpage

$ $
\newpage

$ $
\newpage

$ $
\newpage

$ $
\newpage

$ $
\newpage

$ $
\newpage

$ $
\newpage

$ $
\newpage

$ $
\newpage

$ $
\newpage

$ $
\newpage

$ $
\newpage

\newpage
\begin{thebibliography}{42}
\expandafter\ifx\csname natexlab\endcsname\relax\def\natexlab#1{#1}\fi
\expandafter\ifx\csname bibnamefont\endcsname\relax
  \def\bibnamefont#1{#1}\fi
\expandafter\ifx\csname bibfnamefont\endcsname\relax
  \def\bibfnamefont#1{#1}\fi
\expandafter\ifx\csname citenamefont\endcsname\relax
  \def\citenamefont#1{#1}\fi
\expandafter\ifx\csname url\endcsname\relax
  \def\url#1{\texttt{#1}}\fi
\expandafter\ifx\csname urlprefix\endcsname\relax\def\urlprefix{URL }\fi
\providecommand{\bibinfo}[2]{#2}
\providecommand{\eprint}[2][]{\url{#2}}

\bibitem[{\citenamefont{Doniach}(1977)}]{Doniach:1977}
\bibinfo{author}{\bibfnamefont{S.}~\bibnamefont{Doniach}},
  \bibinfo{journal}{Physica B \& C} \textbf{\bibinfo{volume}{91}},
  \bibinfo{pages}{231 } (\bibinfo{year}{1977}).

\bibitem[{\citenamefont{Steglich et~al.}(1996)\citenamefont{Steglich,
  Buschinger, Gegenwart, Lohmann, Helfrich, Langhammer, Hellmann, Donnevert,
  Thomas, Link et~al.}}]{Steglich:1996}
\bibinfo{author}{\bibfnamefont{F.}~\bibnamefont{Steglich}},
  \bibinfo{author}{\bibfnamefont{B.}~\bibnamefont{Buschinger}},
  \bibinfo{author}{\bibfnamefont{P.}~\bibnamefont{Gegenwart}},
  \bibinfo{author}{\bibfnamefont{M.}~\bibnamefont{Lohmann}},
  \bibinfo{author}{\bibfnamefont{R.}~\bibnamefont{Helfrich}},
  \bibinfo{author}{\bibfnamefont{C.}~\bibnamefont{Langhammer}},
  \bibinfo{author}{\bibfnamefont{P.}~\bibnamefont{Hellmann}},
  \bibinfo{author}{\bibfnamefont{L.}~\bibnamefont{Donnevert}},
  \bibinfo{author}{\bibfnamefont{S.}~\bibnamefont{Thomas}},
  \bibinfo{author}{\bibfnamefont{A.}~\bibnamefont{Link}}, \bibnamefont{et~al.},
  \bibinfo{journal}{J. Phys. Cond. Mat.} \textbf{\bibinfo{volume}{8}},
  \bibinfo{pages}{9909 } (\bibinfo{year}{1996}).

\bibitem[{\citenamefont{Kirkpatrick and Belitz}(2003)}]{Kirkpatrick:2003}
\bibinfo{author}{\bibfnamefont{T.~R.} \bibnamefont{Kirkpatrick}}
  \bibnamefont{and} \bibinfo{author}{\bibfnamefont{D.}~\bibnamefont{Belitz}},
  \bibinfo{journal}{Phys. Rev. B} \textbf{\bibinfo{volume}{67}},
  \bibinfo{pages}{024419} (\bibinfo{year}{2003}).

\bibitem[{\citenamefont{Drescher et~al.}(1996)\citenamefont{Drescher,
  Abd-Elmeguid, Micklitz, and Sanchez}}]{Drescher:1996}
\bibinfo{author}{\bibfnamefont{K.}~\bibnamefont{Drescher}},
  \bibinfo{author}{\bibfnamefont{M.~M.} \bibnamefont{Abd-Elmeguid}},
  \bibinfo{author}{\bibfnamefont{H.}~\bibnamefont{Micklitz}}, \bibnamefont{and}
  \bibinfo{author}{\bibfnamefont{J.~P.} \bibnamefont{Sanchez}},
  \bibinfo{journal}{Phys. Rev. Lett.} \textbf{\bibinfo{volume}{77}},
  \bibinfo{pages}{3228} (\bibinfo{year}{1996}).

\bibitem[{\citenamefont{Drotziger et~al.}(2006)\citenamefont{Drotziger,
  Pfleiderer, Uhlarz, von Lohneysen, Souptel, Loser, and
  Behr}}]{Drotziger:2006}
\bibinfo{author}{\bibfnamefont{S.}~\bibnamefont{Drotziger}},
  \bibinfo{author}{\bibfnamefont{C.}~\bibnamefont{Pfleiderer}},
  \bibinfo{author}{\bibfnamefont{M.}~\bibnamefont{Uhlarz}},
  \bibinfo{author}{\bibfnamefont{H.}~\bibnamefont{v.Lohneysen}},
  \bibinfo{author}{\bibfnamefont{D.}~\bibnamefont{Souptel}},
  \bibinfo{author}{\bibfnamefont{W.}~\bibnamefont{Loser}}, \bibnamefont{and}
  \bibinfo{author}{\bibfnamefont{G.}~\bibnamefont{Behr}},
  \bibinfo{journal}{Phys. Rev. B} \textbf{\bibinfo{volume}{73}},
  \bibinfo{pages}{214413} (\bibinfo{year}{2006}).

\bibitem[{\citenamefont{Sereni et~al.}(2007)\citenamefont{Sereni, Westerkamp,
  Kuchler, Caroca-Canales, Gegenwart, and Geibel}}]{Sereni:2007}
\bibinfo{author}{\bibfnamefont{J.~G.} \bibnamefont{Sereni}},
  \bibinfo{author}{\bibfnamefont{T.}~\bibnamefont{Westerkamp}},
  \bibinfo{author}{\bibfnamefont{R.}~\bibnamefont{Kuchler}},
  \bibinfo{author}{\bibfnamefont{N.}~\bibnamefont{Caroca-Canales}},
  \bibinfo{author}{\bibfnamefont{P.}~\bibnamefont{Gegenwart}},
  \bibnamefont{and} \bibinfo{author}{\bibfnamefont{C.}~\bibnamefont{Geibel}},
  \bibinfo{journal}{Phys. Rev. B} \textbf{\bibinfo{volume}{75}},
  \bibinfo{pages}{024432} (\bibinfo{year}{2007}).

\bibitem[{\citenamefont{Sereni}(1991)}]{Sereni:2006a}
\bibinfo{author}{\bibfnamefont{J.~G.} \bibnamefont{Sereni}}, in
  \emph{\bibinfo{booktitle}{Handbook for Physics and Chemistry of Rare
  Earths}}, edited by \bibinfo{editor}{\bibfnamefont{K.~A.}
  \bibnamefont{Gschneider}} \bibnamefont{and}
  \bibinfo{editor}{\bibfnamefont{L.}~\bibnamefont{Eyring}}
  (\bibinfo{publisher}{North-Holland, Amsterdam}, \bibinfo{year}{1991}),
  vol.~\bibinfo{volume}{XV}, chap.~\bibinfo{chapter}{98}.

\bibitem[{\citenamefont{Zimmer et~al.}(1995)\citenamefont{Zimmer, Jeitschko,
  Albering, Glaum, and Reehuis}}]{Zimmer:1995}
\bibinfo{author}{\bibfnamefont{B.~I.} \bibnamefont{Zimmer}},
  \bibinfo{author}{\bibfnamefont{W.}~\bibnamefont{Jeitschko}},
  \bibinfo{author}{\bibfnamefont{J.~H.} \bibnamefont{Albering}},
  \bibinfo{author}{\bibfnamefont{R.}~\bibnamefont{Glaum}}, \bibnamefont{and}
  \bibinfo{author}{\bibfnamefont{M.}~\bibnamefont{Reehuis}},
  \bibinfo{journal}{J. Alloys Comp.} \textbf{\bibinfo{volume}{229}},
  \bibinfo{pages}{238 } (\bibinfo{year}{1995}).

\bibitem[{\citenamefont{Kaczorowski et~al.}(1994)\citenamefont{Kaczorowski,
  Albering, Noel, and Jeitschko}}]{Kaczorowski:1994}
\bibinfo{author}{\bibfnamefont{D.}~\bibnamefont{Kaczorowski}},
  \bibinfo{author}{\bibfnamefont{J.~H.} \bibnamefont{Albering}},
  \bibinfo{author}{\bibfnamefont{H.}~\bibnamefont{Noel}}, \bibnamefont{and}
  \bibinfo{author}{\bibfnamefont{W.}~\bibnamefont{Jeitschko}},
  \bibinfo{journal}{J. Alloys Comp.} \textbf{\bibinfo{volume}{216}},
  \bibinfo{pages}{117 } (\bibinfo{year}{1994}).

\bibitem[{\citenamefont{Kamihara et~al.}(2006)\citenamefont{Kamihara,
  Hiramatsu, Hirano, Kawamura, Yanagi, Kamiya, and Hosono}}]{Kamihara:2006}
\bibinfo{author}{\bibfnamefont{Y.}~\bibnamefont{Kamihara}},
  \bibinfo{author}{\bibfnamefont{H.}~\bibnamefont{Hiramatsu}},
  \bibinfo{author}{\bibfnamefont{M.}~\bibnamefont{Hirano}},
  \bibinfo{author}{\bibfnamefont{R.}~\bibnamefont{Kawamura}},
  \bibinfo{author}{\bibfnamefont{H.}~\bibnamefont{Yanagi}},
  \bibinfo{author}{\bibfnamefont{T.}~\bibnamefont{Kamiya}}, \bibnamefont{and}
  \bibinfo{author}{\bibfnamefont{H.}~\bibnamefont{Hosono}},
  \bibinfo{journal}{J. Am. Chem. Soc.} \textbf{\bibinfo{volume}{128}},
  \bibinfo{pages}{10012 } (\bibinfo{year}{2006}).

\bibitem[{\citenamefont{Lebegue}(2007)}]{Lebegue:2007}
\bibinfo{author}{\bibfnamefont{S.}~\bibnamefont{Lebegue}},
  \bibinfo{journal}{Phys. Rev. B} \textbf{\bibinfo{volume}{75}},
  \bibinfo{eid}{035110} (\bibinfo{year}{2007}).

\bibitem[{\citenamefont{Kanatzidis et~al.}(2005)\citenamefont{Kanatzidis,
  P\"ottgen, and Jeitschko}}]{Kanatzidis:2005}
\bibinfo{author}{\bibfnamefont{M.~G.} \bibnamefont{Kanatzidis}},
  \bibinfo{author}{\bibfnamefont{R.}~\bibnamefont{P\"ottgen}},
  \bibnamefont{and}
  \bibinfo{author}{\bibfnamefont{W.}~\bibnamefont{Jeitschko}},
  \bibinfo{journal}{Angew. Chemie} \textbf{\bibinfo{volume}{44}},
  \bibinfo{pages}{6996 } (\bibinfo{year}{2005}).

\bibitem[{\citenamefont{Welter et~al.}(2003)\citenamefont{Welter, Morozkin, and
  Halich}}]{Welter:2003}
\bibinfo{author}{\bibfnamefont{R.}~\bibnamefont{Welter}},
  \bibinfo{author}{\bibfnamefont{A.~V.} \bibnamefont{Morozkin}},
  \bibnamefont{and} \bibinfo{author}{\bibfnamefont{K.}~\bibnamefont{Halich}},
  \bibinfo{journal}{J. Magn. Magn. Mater.} \textbf{\bibinfo{volume}{257}},
  \bibinfo{pages}{44 } (\bibinfo{year}{2003}).

\bibitem[{\citenamefont{Morozkin et~al.}(2005)\citenamefont{Morozkin, Halich,
  Welter, and Ouladdiaf}}]{Morozkin:2005}
\bibinfo{author}{\bibfnamefont{A.}~\bibnamefont{Morozkin}},
  \bibinfo{author}{\bibfnamefont{K.}~\bibnamefont{Halich}},
  \bibinfo{author}{\bibfnamefont{R.}~\bibnamefont{Welter}}, \bibnamefont{and}
  \bibinfo{author}{\bibfnamefont{B.}~\bibnamefont{Ouladdiaf}},
  \bibinfo{journal}{J. Alloys Compd.} \textbf{\bibinfo{volume}{393}},
  \bibinfo{pages}{34 } (\bibinfo{year}{2005}).

\bibitem[{\citenamefont{Besnus et~al.}(1992)\citenamefont{Besnus, Braghta,
  Hamdaoui, and Meyer}}]{Besnus:1992}
\bibinfo{author}{\bibfnamefont{M.~J.} \bibnamefont{Besnus}},
  \bibinfo{author}{\bibfnamefont{A.}~\bibnamefont{Braghta}},
  \bibinfo{author}{\bibfnamefont{N.}~\bibnamefont{Hamdaoui}}, \bibnamefont{and}
  \bibinfo{author}{\bibfnamefont{A.}~\bibnamefont{Meyer}}, \bibinfo{journal}{J.
  Magn. Magn. Mater.} \textbf{\bibinfo{volume}{104}}, \bibinfo{pages}{1385 }
  (\bibinfo{year}{1992}).

\bibitem[{\citenamefont{Campoy et~al.}(2006)\citenamefont{Campoy, Plaza,
  Coelho, and Gama}}]{Campoy:2006}
\bibinfo{author}{\bibfnamefont{J.~C.~P.} \bibnamefont{Campoy}},
  \bibinfo{author}{\bibfnamefont{E.~J.~R.} \bibnamefont{Plaza}},
  \bibinfo{author}{\bibfnamefont{A.~A.} \bibnamefont{Coelho}},
  \bibnamefont{and} \bibinfo{author}{\bibfnamefont{S.}~\bibnamefont{Gama}},
  \bibinfo{journal}{Phys. Rev. B} \textbf{\bibinfo{volume}{74}},
  \bibinfo{pages}{134410} (\bibinfo{year}{2006}).

\bibitem[{\citenamefont{Garde and Ray}(1995)}]{Garde:1995}
\bibinfo{author}{\bibfnamefont{C.~S.} \bibnamefont{Garde}} \bibnamefont{and}
  \bibinfo{author}{\bibfnamefont{J.}~\bibnamefont{Ray}},
  \bibinfo{journal}{Phys. Rev. B} \textbf{\bibinfo{volume}{51}},
  \bibinfo{pages}{2960} (\bibinfo{year}{1995}).

\bibitem[{\citenamefont{Amato et~al.}(1989)\citenamefont{Amato, Jaccard,
  Sierro, Haen, Lejay, and Flouquet}}]{Amato:1989}
\bibinfo{author}{\bibfnamefont{A.}~\bibnamefont{Amato}},
  \bibinfo{author}{\bibfnamefont{D.}~\bibnamefont{Jaccard}},
  \bibinfo{author}{\bibfnamefont{J.}~\bibnamefont{Sierro}},
  \bibinfo{author}{\bibfnamefont{P.}~\bibnamefont{Haen}},
  \bibinfo{author}{\bibfnamefont{P.}~\bibnamefont{Lejay}}, \bibnamefont{and}
  \bibinfo{author}{\bibfnamefont{J.}~\bibnamefont{Flouquet}},
  \bibinfo{journal}{J. Low. Temp. Phys.} \textbf{\bibinfo{volume}{77}},
  \bibinfo{pages}{195 } (\bibinfo{year}{1989}).

\bibitem[{\citenamefont{Zlati\ifmmode~\acute{c}\else \'{c}\fi{} and
  Monnier}(2005)}]{Zlatic:2005}
\bibinfo{author}{\bibfnamefont{V.}~\bibnamefont{Zlati\ifmmode~\acute{c}\else
  \'{c}\fi{}}} \bibnamefont{and}
  \bibinfo{author}{\bibfnamefont{R.}~\bibnamefont{Monnier}},
  \bibinfo{journal}{Phys. Rev. B} \textbf{\bibinfo{volume}{71}},
  \bibinfo{pages}{165109} (\bibinfo{year}{2005}).

\bibitem[{\citenamefont{Wilhelm and Jaccard}(2004)}]{Wilhelm:2004}
\bibinfo{author}{\bibfnamefont{H.}~\bibnamefont{Wilhelm}} \bibnamefont{and}
  \bibinfo{author}{\bibfnamefont{D.}~\bibnamefont{Jaccard}},
  \bibinfo{journal}{Phys. Rev. B} \textbf{\bibinfo{volume}{69}},
  \bibinfo{pages}{214408} (\bibinfo{year}{2004}).

\bibitem[{\citenamefont{Houshiar et~al.}(1995)\citenamefont{Houshiar, Adroja,
  and Rainford}}]{Houshiar:1995}
\bibinfo{author}{\bibfnamefont{M.}~\bibnamefont{Houshiar}},
  \bibinfo{author}{\bibfnamefont{D.~T.} \bibnamefont{Adroja}},
  \bibnamefont{and} \bibinfo{author}{\bibfnamefont{B.~D.}
  \bibnamefont{Rainford}}, \bibinfo{journal}{J. Magn. Magn. Mater.}
  \textbf{\bibinfo{volume}{140}}, \bibinfo{pages}{1231 }
  (\bibinfo{year}{1995}).

\bibitem[{\citenamefont{Zlatic et~al.}(2003)\citenamefont{Zlatic, Horvatic,
  Milat, Coqblin, Czycholl, and Grenzebach}}]{Zlatic:2003}
\bibinfo{author}{\bibfnamefont{V.}~\bibnamefont{Zlatic}},
  \bibinfo{author}{\bibfnamefont{B.}~\bibnamefont{Horvatic}},
  \bibinfo{author}{\bibfnamefont{I.}~\bibnamefont{Milat}},
  \bibinfo{author}{\bibfnamefont{B.}~\bibnamefont{Coqblin}},
  \bibinfo{author}{\bibfnamefont{G.}~\bibnamefont{Czycholl}}, \bibnamefont{and}
  \bibinfo{author}{\bibfnamefont{C.}~\bibnamefont{Grenzebach}},
  \bibinfo{journal}{Phys. Rev. B} \textbf{\bibinfo{volume}{68}},
  \bibinfo{pages}{104432} (\bibinfo{year}{2003}).

\bibitem[{\citenamefont{Hartmann et~al.}(2006)\citenamefont{Hartmann, Kohler,
  Oeschler, Paschen, Krellner, Geibel, and Steglich}}]{Hartmann:2006}
\bibinfo{author}{\bibfnamefont{S.}~\bibnamefont{Hartmann}},
  \bibinfo{author}{\bibfnamefont{U.}~\bibnamefont{Kohler}},
  \bibinfo{author}{\bibfnamefont{N.}~\bibnamefont{Oeschler}},
  \bibinfo{author}{\bibfnamefont{S.}~\bibnamefont{Paschen}},
  \bibinfo{author}{\bibfnamefont{C.}~\bibnamefont{Krellner}},
  \bibinfo{author}{\bibfnamefont{C.}~\bibnamefont{Geibel}}, \bibnamefont{and}
  \bibinfo{author}{\bibfnamefont{F.}~\bibnamefont{Steglich}},
  \bibinfo{journal}{Physica B} \textbf{\bibinfo{volume}{378-80}},
  \bibinfo{pages}{70 } (\bibinfo{year}{2006}).

\bibitem[{\citenamefont{Larrea et~al.}(2005)\citenamefont{Larrea, Fontes,
  Alvarenga, Baggio-Saitovitch, Burghardt, Eichler, and
  Continentino}}]{Larrea:2005}
\bibinfo{author}{\bibfnamefont{J.} \bibnamefont{Larrea J.}},
  \bibinfo{author}{\bibfnamefont{M.~B.} \bibnamefont{Fontes}},
  \bibinfo{author}{\bibfnamefont{A.~D.} \bibnamefont{Alvarenga}},
  \bibinfo{author}{\bibfnamefont{E.~M.} \bibnamefont{Baggio-Saitovitch}},
  \bibinfo{author}{\bibfnamefont{T.}~\bibnamefont{Burghardt}},
  \bibinfo{author}{\bibfnamefont{A.}~\bibnamefont{Eichler}}, \bibnamefont{and}
  \bibinfo{author}{\bibfnamefont{M.~A.} \bibnamefont{Continentino}},
  \bibinfo{journal}{Phys. Rev. B} \textbf{\bibinfo{volume}{72}},
  \bibinfo{pages}{035129} (\bibinfo{year}{2005}).

\bibitem[{\citenamefont{Carter and Bennett}(1977)}]{Carter:1977}
\bibinfo{author}{\bibfnamefont{G.~C.} \bibnamefont{Carter}} \bibnamefont{and}
  \bibinfo{author}{\bibfnamefont{L.}~\bibnamefont{Bennett}},
  \emph{\bibinfo{title}{Metallic Shifts in NMR}} (\bibinfo{publisher}{Pergamon
  Press}, \bibinfo{year}{1977}).

\bibitem[{\citenamefont{Kobayashi et~al.}(1996)\citenamefont{Kobayashi, Takagi,
  Uesawa, Haga, and Suzuki}}]{Kobayashi:1996}
\bibinfo{author}{\bibfnamefont{A.}~\bibnamefont{Kobayashi}},
  \bibinfo{author}{\bibfnamefont{S.}~\bibnamefont{Takagi}},
  \bibinfo{author}{\bibfnamefont{A.}~\bibnamefont{Uesawa}},
  \bibinfo{author}{\bibfnamefont{Y.}~\bibnamefont{Haga}}, \bibnamefont{and}
  \bibinfo{author}{\bibfnamefont{T.}~\bibnamefont{Suzuki}},
  \bibinfo{journal}{J. Phys. Soc. Jpn.} \textbf{\bibinfo{volume}{65}},
  \bibinfo{pages}{3343 } (\bibinfo{year}{1996}).

\bibitem[{Bru()}]{Bruening}
\bibinfo{note}{E. M. Br\"uning (unpublished)}.

\bibitem[{\citenamefont{Nakamura et~al.}(1996)\citenamefont{Nakamura,
  Takayanagi, Shiga, Nishi, and Kakurai}}]{Nakamura:1996}
\bibinfo{author}{\bibfnamefont{H.}~\bibnamefont{Nakamura}},
  \bibinfo{author}{\bibfnamefont{F.}~\bibnamefont{Takayanagi}},
  \bibinfo{author}{\bibfnamefont{M.}~\bibnamefont{Shiga}},
  \bibinfo{author}{\bibfnamefont{M.}~\bibnamefont{Nishi}}, \bibnamefont{and}
  \bibinfo{author}{\bibfnamefont{K.}~\bibnamefont{Kakurai}},
  \bibinfo{journal}{J. Phys. Soc. Jpn.} \textbf{\bibinfo{volume}{65}},
  \bibinfo{pages}{2779 } (\bibinfo{year}{1996}).

\bibitem[{\citenamefont{B\"uttgen et~al.}(1996)\citenamefont{B\"uttgen, Bohmer,
  Krimmel, and Loidl}}]{Buttgen:1996}
\bibinfo{author}{\bibfnamefont{N.}~\bibnamefont{B\"uttgen}},
  \bibinfo{author}{\bibfnamefont{R.}~\bibnamefont{Bohmer}},
  \bibinfo{author}{\bibfnamefont{A.}~\bibnamefont{Krimmel}}, \bibnamefont{and}
  \bibinfo{author}{\bibfnamefont{A.}~\bibnamefont{Loidl}},
  \bibinfo{journal}{Phys. Rev. B} \textbf{\bibinfo{volume}{53}},
  \bibinfo{pages}{5557 } (\bibinfo{year}{1996}).

\bibitem[{\citenamefont{Koepernik and Eschrig}(1999)}]{FPLO}
\bibinfo{author}{\bibfnamefont{K.}~\bibnamefont{Koepernik}} \bibnamefont{and}
  \bibinfo{author}{\bibfnamefont{H.}~\bibnamefont{Eschrig}},
  \bibinfo{journal}{Phys. Rev. B} \textbf{\bibinfo{volume}{59}},
  \bibinfo{pages}{1743} (\bibinfo{year}{1999}).

\bibitem[{\citenamefont{Perdew and Wang}(1992)}]{PW}
\bibinfo{author}{\bibfnamefont{J.~P.} \bibnamefont{Perdew}} \bibnamefont{and}
  \bibinfo{author}{\bibfnamefont{Y.}~\bibnamefont{Wang}},
  \bibinfo{journal}{Phys. Rev. B} \textbf{\bibinfo{volume}{45}},
  \bibinfo{pages}{13244} (\bibinfo{year}{1992}).

\bibitem[{\citenamefont{Eschrig}(1989)}]{HE}
\bibinfo{author}{\bibfnamefont{H.}~\bibnamefont{Eschrig}},
  \emph{\bibinfo{title}{Optimized LCAO Method and the Electronic Structure of
  Extended Systems}} (\bibinfo{publisher}{Springer, Berlin},
  \bibinfo{year}{1989}).

\bibitem[{\citenamefont{Shick et~al.}(2001)\citenamefont{Shick, Pickett, and
  Liechtenstein}}]{SP}
\bibinfo{author}{\bibfnamefont{A.~B.} \bibnamefont{Shick}},
  \bibinfo{author}{\bibfnamefont{W.~E.} \bibnamefont{Pickett}},
  \bibnamefont{and} \bibinfo{author}{\bibfnamefont{A.~I.}
  \bibnamefont{Liechtenstein}}, \bibinfo{journal}{Journal of Electron
  Spectroscopy and Related Phenomena} \textbf{\bibinfo{volume}{114-116}},
  \bibinfo{pages}{753} (\bibinfo{year}{2001}).

\bibitem[{\citenamefont{Blaha et~al.}(2001)\citenamefont{Blaha, Schwarz,
  Madsen, Kvasnicka, and Luitz}}]{WIEN}
\bibinfo{author}{\bibfnamefont{P.}~\bibnamefont{Blaha}},
  \bibinfo{author}{\bibfnamefont{K.}~\bibnamefont{Schwarz}},
  \bibinfo{author}{\bibfnamefont{G.}~\bibnamefont{Madsen}},
  \bibinfo{author}{\bibfnamefont{D.}~\bibnamefont{Kvasnicka}},
  \bibnamefont{and} \bibinfo{author}{\bibfnamefont{J.}~\bibnamefont{Luitz}},
  \emph{\bibinfo{title}{WIEN2k, An Augmented Plane Wave Local Orbitals Program
  for Calculating Crystal Properties}} (\bibinfo{publisher}{Karlheinz Schwarz,
  Techn. Universitadt Wien, Austria}, \bibinfo{year}{2001}), ISBN
  \bibinfo{isbn}{3-9501031-1-2}.

\bibitem[{Ros()}]{Rosner}
\bibinfo{note}{546 (576) $k$-points in the irreducible part of the BZ for the
  simple cell (super cell). In both cases: $R_{MT}(\tn{Ce})=2.34$\,au,
  $R_{MT}(\tn{Ru})=2.32$\,au, $R_{MT}(\tn{P})=2.06$\,au and
  $R_{MT}(\tn{O})=2.08$\,au. $R_{MT}K_{\tn{max}}=7.0$, LSDA+$U$ with separation
  energy\,=\,-6.0\,Ry, $U=0.45$\,Ry and $J=0.05$\,Ry.}

\bibitem[{\citenamefont{Chevalier and Malaman}(2004)}]{Chevalier:2004}
\bibinfo{author}{\bibfnamefont{B.}~\bibnamefont{Chevalier}} \bibnamefont{and}
  \bibinfo{author}{\bibfnamefont{B.}~\bibnamefont{Malaman}},
  \bibinfo{journal}{Solid State Commun.} \textbf{\bibinfo{volume}{130}},
  \bibinfo{pages}{711 } (\bibinfo{year}{2004}).

\bibitem[{\citenamefont{Skorek et~al.}(2001)\citenamefont{Skorek, Deniszczyk,
  Szade, and Tyszka}}]{Skorek:2001}
\bibinfo{author}{\bibfnamefont{G.}~\bibnamefont{Skorek}},
  \bibinfo{author}{\bibfnamefont{J.}~\bibnamefont{Deniszczyk}},
  \bibinfo{author}{\bibfnamefont{J.}~\bibnamefont{Szade}}, \bibnamefont{and}
  \bibinfo{author}{\bibfnamefont{B.}~\bibnamefont{Tyszka}},
  \bibinfo{journal}{J. Phys. C} \textbf{\bibinfo{volume}{13}},
  \bibinfo{pages}{6397 } (\bibinfo{year}{2001}).

\bibitem[{\citenamefont{Sidorov et~al.}(2003)\citenamefont{Sidorov, Bauer,
  Frederick, Jeffries, Nakatsuji, Moreno, Thompson, Maple, and
  Fisk}}]{Sidorov:2003}
\bibinfo{author}{\bibfnamefont{V.~A.} \bibnamefont{Sidorov}},
  \bibinfo{author}{\bibfnamefont{E.~D.} \bibnamefont{Bauer}},
  \bibinfo{author}{\bibfnamefont{N.~A.} \bibnamefont{Frederick}},
  \bibinfo{author}{\bibfnamefont{J.~R.} \bibnamefont{Jeffries}},
  \bibinfo{author}{\bibfnamefont{S.}~\bibnamefont{Nakatsuji}},
  \bibinfo{author}{\bibfnamefont{N.~O.}~\bibnamefont{Moreno}},
  \bibinfo{author}{\bibfnamefont{J.~D.} \bibnamefont{Thompson}},
  \bibinfo{author}{\bibfnamefont{M.~B.} \bibnamefont{Maple}}, \bibnamefont{and}
  \bibinfo{author}{\bibfnamefont{Z.}~\bibnamefont{Fisk}},
  \bibinfo{journal}{Phys. Rev. B} \textbf{\bibinfo{volume}{67}},
  \bibinfo{pages}{224419} (\bibinfo{year}{2003}).

\bibitem[{\citenamefont{Pierre et~al.}(1990)\citenamefont{Pierre, Laborde,
  Houssay, Rouault, Senateur, and Madar}}]{Pierre:1990}
\bibinfo{author}{\bibfnamefont{J.}~\bibnamefont{Pierre}},
  \bibinfo{author}{\bibfnamefont{O.}~\bibnamefont{Laborde}},
  \bibinfo{author}{\bibfnamefont{E.}~\bibnamefont{Houssay}},
  \bibinfo{author}{\bibfnamefont{A.}~\bibnamefont{Rouault}},
  \bibinfo{author}{\bibfnamefont{J.~P.} \bibnamefont{Senateur}},
  \bibnamefont{and} \bibinfo{author}{\bibfnamefont{R.}~\bibnamefont{Madar}},
  \bibinfo{journal}{J. Phys. C} \textbf{\bibinfo{volume}{2}},
  \bibinfo{pages}{431 } (\bibinfo{year}{1990}).

\bibitem[{\citenamefont{Koelling et~al.}(1985)\citenamefont{Koelling, Dunlap,
  and Crabtree}}]{Koelling:1985}
\bibinfo{author}{\bibfnamefont{D.~D.} \bibnamefont{Koelling}},
  \bibinfo{author}{\bibfnamefont{B.~D.} \bibnamefont{Dunlap}},
  \bibnamefont{and} \bibinfo{author}{\bibfnamefont{G.~W.}
  \bibnamefont{Crabtree}}, \bibinfo{journal}{Phys. Rev. B}
  \textbf{\bibinfo{volume}{31}}, \bibinfo{pages}{4966 } (\bibinfo{year}{1985}).

\bibitem[{\citenamefont{Petrovic et~al.}(2001)\citenamefont{Petrovic, Pagliuso,
  Hundley, Movshovich, Sarrao, Thompson, Fisk, and Monthoux}}]{Petrovic:2001}
\bibinfo{author}{\bibfnamefont{C.}~\bibnamefont{Petrovic}},
  \bibinfo{author}{\bibfnamefont{P.~G.} \bibnamefont{Pagliuso}},
  \bibinfo{author}{\bibfnamefont{M.~F.} \bibnamefont{Hundley}},
  \bibinfo{author}{\bibfnamefont{R.}~\bibnamefont{Movshovich}},
  \bibinfo{author}{\bibfnamefont{J.~L.} \bibnamefont{Sarrao}},
  \bibinfo{author}{\bibfnamefont{J.~D.} \bibnamefont{Thompson}},
  \bibinfo{author}{\bibfnamefont{Z.}~\bibnamefont{Fisk}}, \bibnamefont{and}
  \bibinfo{author}{\bibfnamefont{P.}~\bibnamefont{Monthoux}},
  \bibinfo{journal}{J. Phys. C} \textbf{\bibinfo{volume}{13}},
  \bibinfo{pages}{L337 } (\bibinfo{year}{2001}).

\bibitem[{\citenamefont{Settai et~al.}(2001)\citenamefont{Settai, Shishido,
  Ikeda, Murakawa, Nakashima, Aoki, Haga, Harima, and Onuki}}]{Settai:2001}
\bibinfo{author}{\bibfnamefont{R.}~\bibnamefont{Settai}},
  \bibinfo{author}{\bibfnamefont{H.}~\bibnamefont{Shishido}},
  \bibinfo{author}{\bibfnamefont{S.}~\bibnamefont{Ikeda}},
  \bibinfo{author}{\bibfnamefont{Y.}~\bibnamefont{Murakawa}},
  \bibinfo{author}{\bibfnamefont{M.}~\bibnamefont{Nakashima}},
  \bibinfo{author}{\bibfnamefont{D.}~\bibnamefont{Aoki}},
  \bibinfo{author}{\bibfnamefont{Y.}~\bibnamefont{Haga}},
  \bibinfo{author}{\bibfnamefont{H.}~\bibnamefont{Harima}}, \bibnamefont{and}
  \bibinfo{author}{\bibfnamefont{Y.}~\bibnamefont{Onuki}}, \bibinfo{journal}{J.
  Phys. C} \textbf{\bibinfo{volume}{13}}, \bibinfo{pages}{L627 }
  (\bibinfo{year}{2001}).

\end{thebibliography}
\end{document}